\documentclass{PHYEAUTH}
\usepackage{graphicx}
\usepackage{amsmath}
\usepackage{amssymb}
\input{epsf}

\begin{document}

\begin{frontmatter}

\title{High-frequency blockade and related phenomena}

\author[address1,address2]{M.M. Mahmoodian\thanksref{thank1}},
\author[address1]{M.V. Entin} and
\author[address1]{L.S. Braginsky}

\address[address1]{Institute of Semiconductor Physics, Siberian
Division of Russian Academy of Sciences, Novosibirsk, 630090,
Russia}

\address[address2]{Novosibirsk State University, Novosibirsk, 630090,
Russia}

\thanks[thank1]{Corresponding author. E-mail: mahmood@isp.nsc.ru}

\begin{abstract}
We study systems with local vibrating potentials, one-dimensional
single and double wells and the tight-binding 1D model with single
vibrating site. In general, these systems transmit, or reflect
particles inelastically, with absorption or emission of several
frequency quanta. Nevertheless, we have found that at some
conditions these systems can perfectly and elastically reflect
electrons. This high-frequency "blockade" give rise to unique
possibility of near-ideal localization of electron with the energy
lying on the background of continuous energy spectrum. We discuss
different consequences of this statement.
\end{abstract}

\end{frontmatter}

\vspace{0.1 cm}\noindent{\bf Basic formulae.}\vspace{0.1 cm}

We consider 1D systems with potential
\begin{eqnarray}\label{01}
U(x)=(u_1+v_1\sin(\omega t))\delta(x+d)+\nonumber\\
(u_2+v_2\sin(\omega t+\varphi))\delta(x-d),
\end{eqnarray} where $t$ is time, $\hbar=m_e=1$, the case of
single well is specified by $u_2=v_2=0$), and the tight-binding
model
\begin{equation}\label{a}
i\dot{a}_m-\delta_{m,0}(\widetilde{u}+\widetilde{v}\sin\omega
t)a_0+\frac{\Delta}{2}(a_{m+1}+a_{m-1})=0.
\end{equation}
Here $a_m$ is time-dependent amplitude of the wave function on the
$m$-th atom,  $\widetilde{u}+\widetilde{v}\sin\omega t$ is the
vibrating energy level of the atom $m=0$, $2\Delta$ is the width
of permitted band, the  energy is referred to the band center. The
presence of alternating potential leads to change of the dc
conductance and appearance of stationary current excited by
alternating potential itself (electron pumping
\cite{Thouless,Brouwer,Moskalets,brag,mah,mah1}). The present
paper consider these responses beyond the perturbation theory
approximation, that leads to new unusual features.

The electron wave functions outside the wells (barriers) can be
searched in the form:
\begin{eqnarray}\label{wf}
&&\Psi=e^{-iEt}\Big\{e^{ipx}+\sum\limits_n r_ne^{-in\omega t-ip_nx}\Big\}~~~\mbox{ left,}\\
&&\Psi=e^{-iEt}\sum\limits_n e^{-in\omega t+ip_nx}t_n
~~~~~~~~~~~~~~~~~\mbox{ right.}
\end{eqnarray}
Here $p_n=\sqrt{p^2+2n\omega},~p=\sqrt{2E}$. In the single-well
problem case the transmission coefficients $t_n$ obeys
\begin{eqnarray}\label{single}
t_n(p_n+iu)+\frac{v}{2}(t_{n+1}-t_{n-1})=p_n\delta_{n,0}.
\end{eqnarray}
The Eq. (\ref{single}) can be found from equations for
corresponding double-barrier problem \cite{brag,mah}. At low
temperature the conductance $G$, the stationary current $J_0$,
caused by alternating signal itself, and derivative of $J_0$ with
respect to the Fermi level $E_F$ read:
\begin{eqnarray}\label{G}
G=\frac{G_0}{2}\left(T^\rightarrow+T^\leftarrow\right)|_{E=E_F},\\\label{J}
J_0=\frac{e}{\pi\hbar}\int_0^{E_F}dE\left(T^\rightarrow-T^\leftarrow\right),\\
e\frac{\partial}{\partial E_F}J_0\equiv
S=G_0\left(T^\rightarrow-T^\leftarrow\right)|_{E=E_F}.\label{S}
\end{eqnarray}
Here arrows mark the direction of motion, $T^\rightleftarrows
=\sum_n|t_n^\rightleftarrows|^2$, $G_0=2e^2/h$ is the conductance
quantum. The Eq. (\ref{G}) results from Landauer formula, the Eqs.
(\ref{J}-\ref{S}) see in  \cite{brag,mah}. The stationary current
$J_0$ exists if the system is asymmetric, $T^\rightarrow\neq
T^\leftarrow$, in particular, it vanishes for single
$\delta$-function and the model (\ref{a}).

\vspace{0.1 cm}\noindent{\bf Single oscillating $\delta$-function.
Conductance and reflection resonances.}\vspace{0.1 cm}

\begin{figure}[ht]
\centerline{\epsfysize=2.7cm\epsfbox{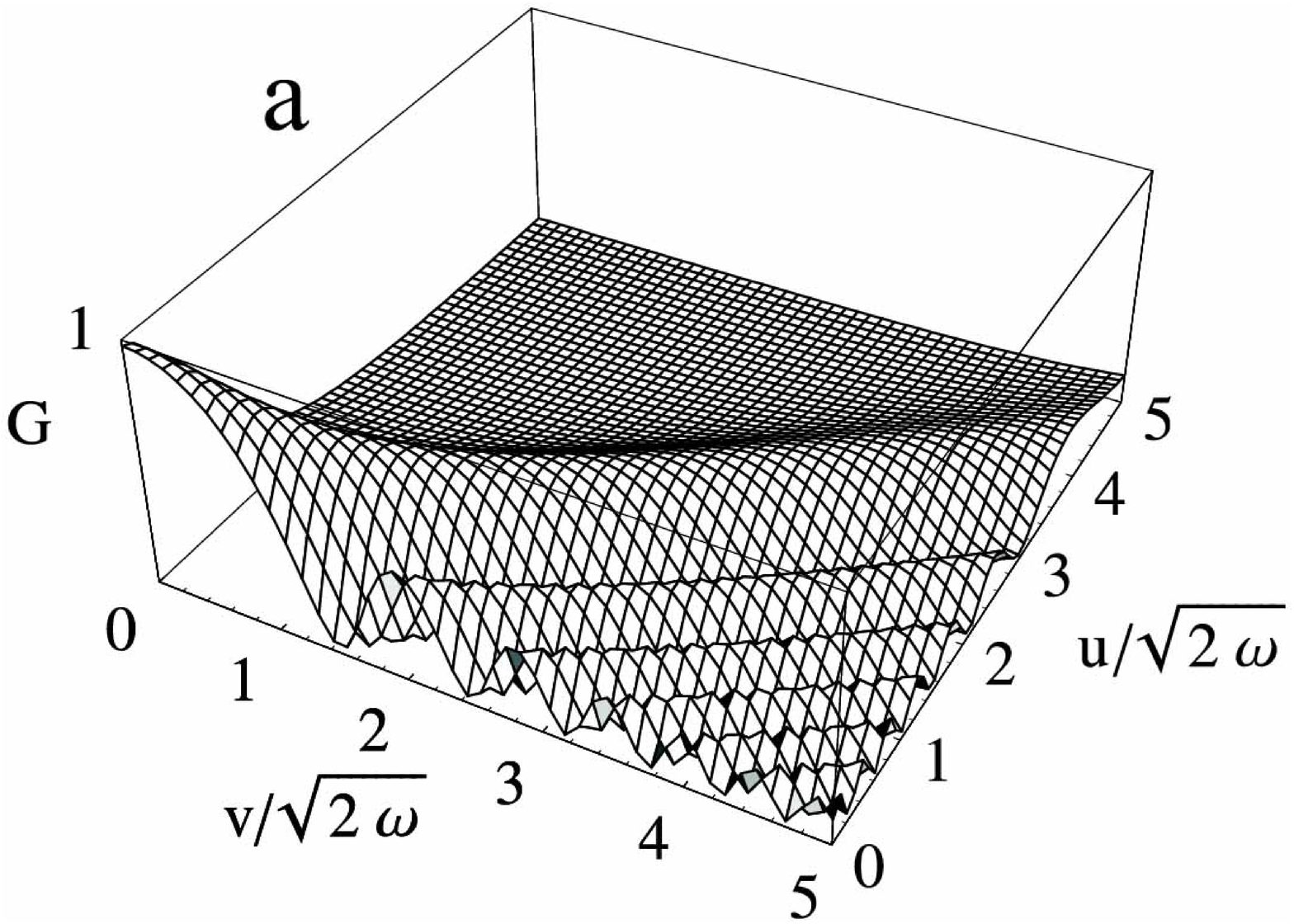}
\epsfysize=2.7cm\epsfbox{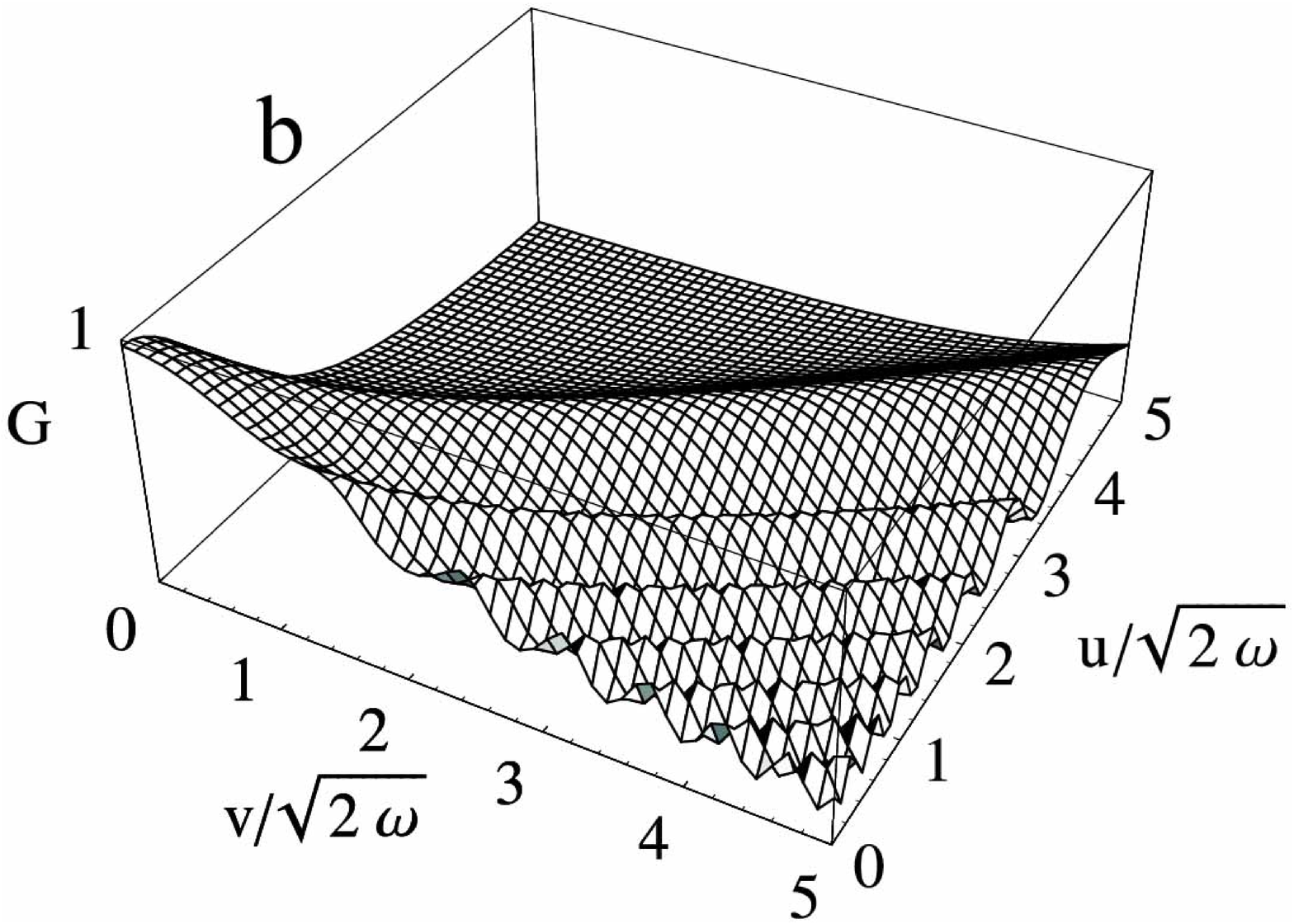}}
\centerline{\epsfysize=2cm\epsfbox{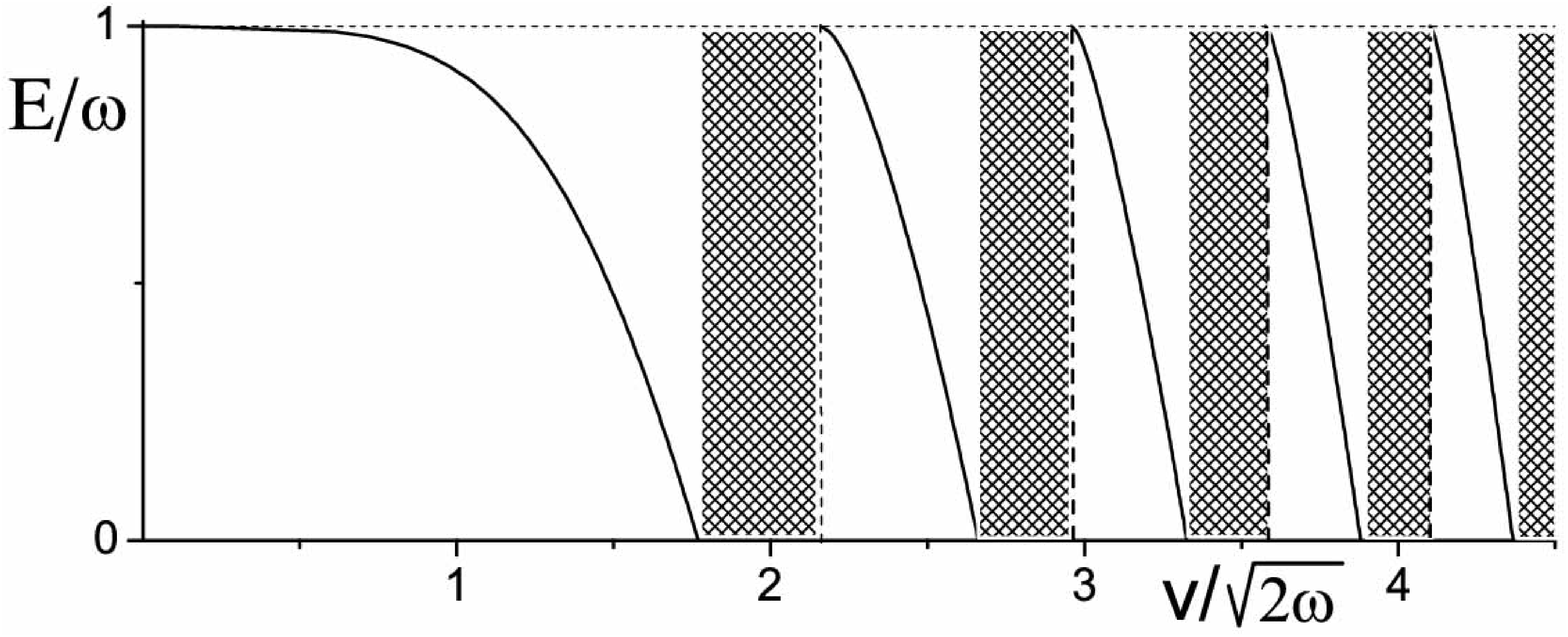}} \caption{a,b -
Conductance of single vibrating barrier/well in units of $G_0$
versus $u$ and $v$. The Fermi energy has values $ 0.5\omega$ (a)
and $1.5\omega$ (b). c - Blockade state energy versus value of
alternating potential $v$ for $u=0$. Forbidden values of $v$,
where blockade states absent are crosshatched.}\label{cond}
\end{figure}

Figure~\ref{cond} presents dependence of conductance on parameters
of the potential.
Apparently, the conductance is oscillating function of the
amplitude of vibrations $v$. The conductance vanishes at the lines
of the $(v,u)$ plane, if $E_F<\omega$; otherwise, if $E_F>\omega$,
it oscillates remaining positive.

Zeros of the conductance coincides with  zeros of the
transmittance. It is important that not only transmission, but
also inelasticity disappear at the resonance. We call this
phenomenon "high-frequency blockade" by analogy with the "Coulomb
blockade". Unlike the latter, the "high-frequency blockade" occurs
in the absence of interaction between the electrons.

Figure~\ref{cond}c presents dependence of the blockade eigenvalues
in the case $u=0$, found by solving the homogeneous system of
equations (\ref{single}) for $n\leq -1$, while for $n\geq 0$
$t_n=0$. The blockade states exist only for $E<\omega$. Only one
reflection resonance occurs under small $v$.  Note, that zeros of
the conductance disappears at some intervals of $v$
(Fig.~\ref{cond}c).

\begin{figure}[ht]
\centerline{\epsfysize=3.5cm\epsfbox{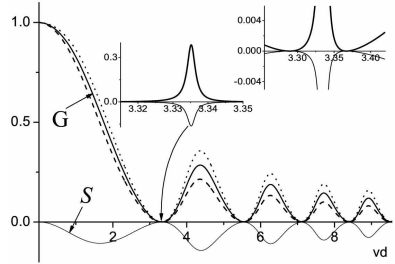}}
\caption{Conductance (bold) and derivative of photocurrent (thin)
in double-well structure versus amplitude of vibrations. The
problem parameters:
$\omega=5,~p_F=\pi/2d,~\varphi=\pi/2,~u_1=u_2=0,~v=v_1=v_2$. The
inserts show magnified vicinity of conductance zeros.}\label{sym}
\end{figure}

\vspace{0.1 cm}\noindent{\bf Two oscillating
$\delta-$functions.}\vspace{0.1 cm}

This system represents the electron "Fabri-Perrote" interferometer
with vibrating mirrors. Due to asymmetry, it can serve as a
quantum pump. Figure~\ref{sym} presents dependence of conductance
$G$ and the chemical potential derivative of the photocurrent $S$
on $v=v_1=v_2$ for $u=0$. The distance between $\delta$-functions
is chosen equal to half of the Fermi wavelength. Roughly the
dependence resembles that for the case of one $\delta-$function
(Fig.~\ref{cond}a). Both values $G$ and $S$ vanish near the same
points. However, strong magnification shows that zeros are
splitted squeezing very narrow transmission resonances. They arise
from quasilocal states confined between $\delta$-functions. Widths
of these states are determined by mixing of exponentially decaying
modes with imaginary $p_n$ for $E+n\omega<0$; the quasilocal
states are drastically narrowing when the distance between
$\delta$-functions grows.

\vspace{0.1 cm}\noindent{\bf 1D tight-binding model with one
oscillating atom.}\vspace{0.1 cm}

In this case the Eq. (\ref{single}) stays valid with replacement
$p_n\to \sqrt{(\Delta^2-(E+n\omega)^2)/\Delta},
u\to\widetilde{u}/\sqrt{\Delta}, v\to\widetilde{v}/\sqrt{\Delta}$.

\begin{figure}[ht]
\centerline{\epsfysize=3.7cm\epsfbox{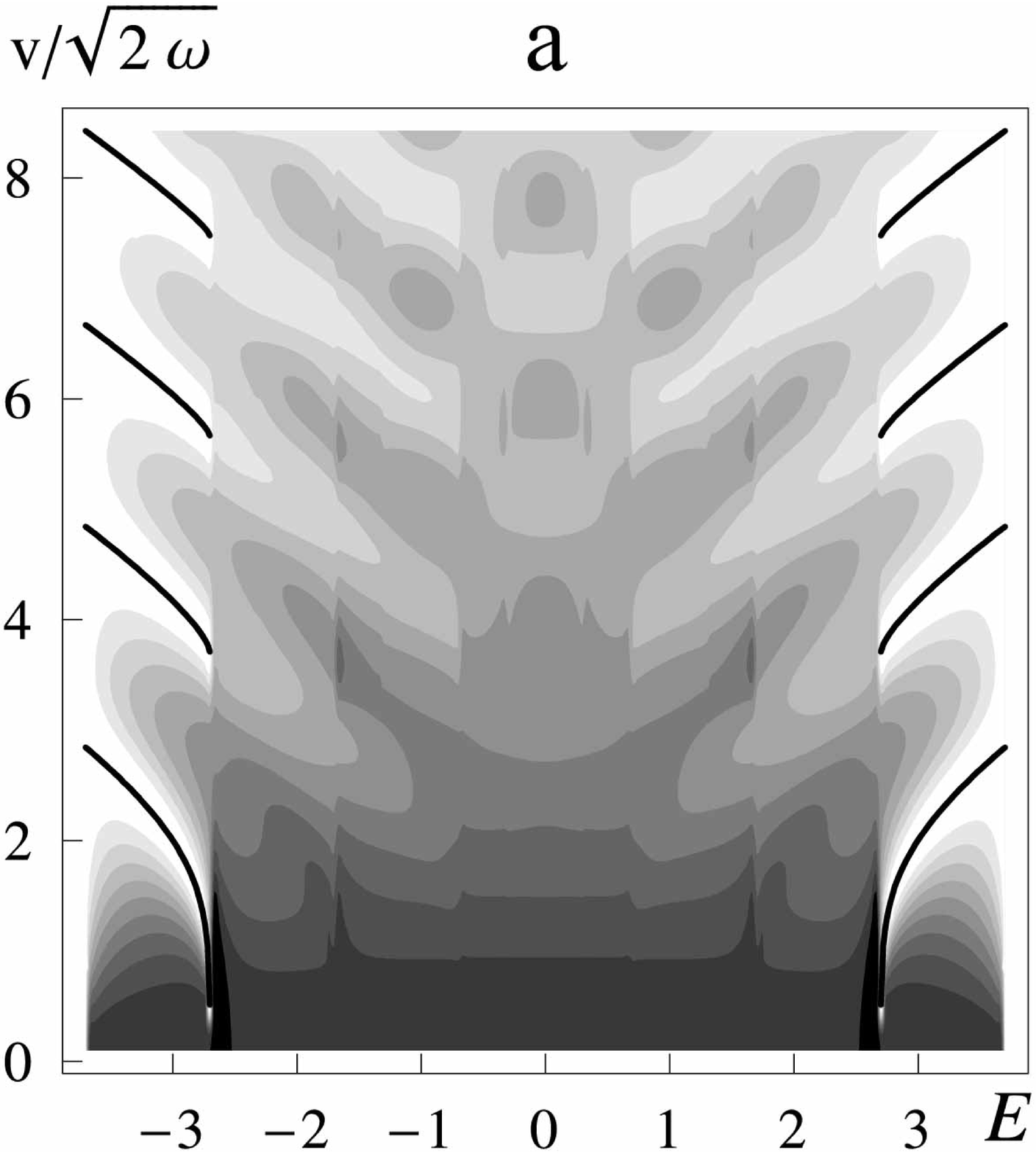}\hspace{0.2cm}\epsfysize=2.5cm\epsfbox{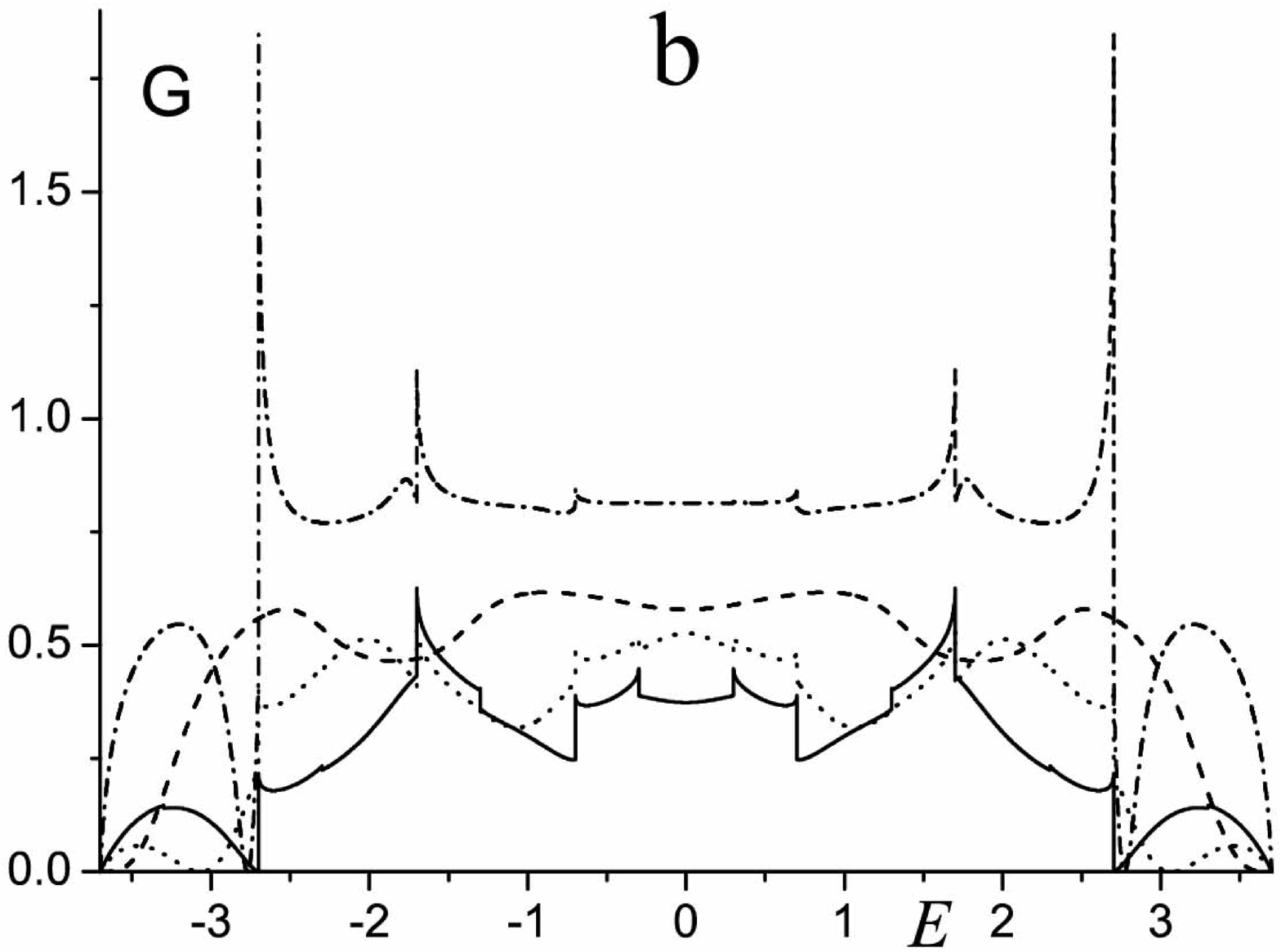}}\vspace{0.1cm}
\centerline{\epsfysize=3.7cm\epsfbox{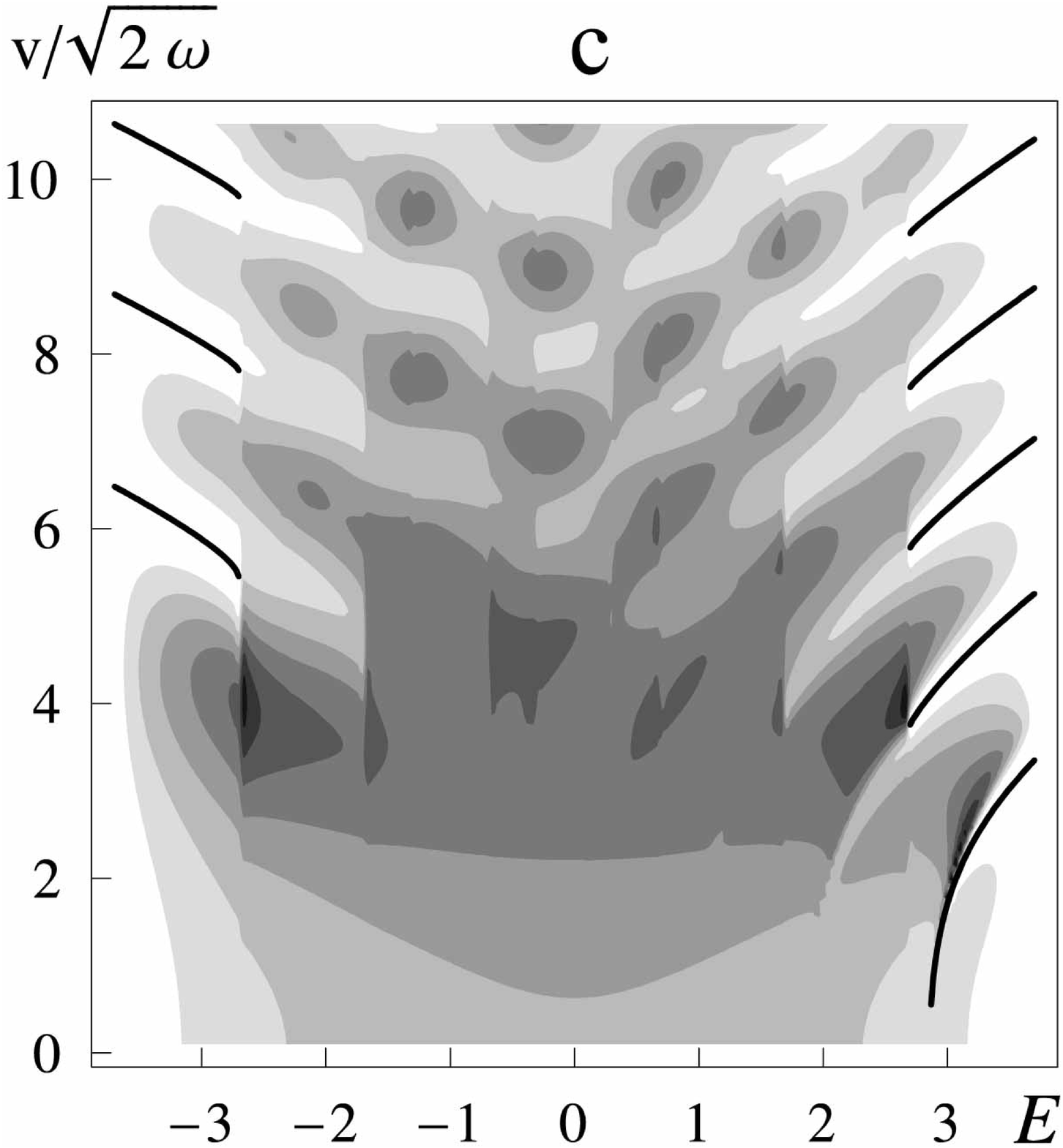}\hspace{0.2cm}\epsfysize=3.7cm\epsfbox{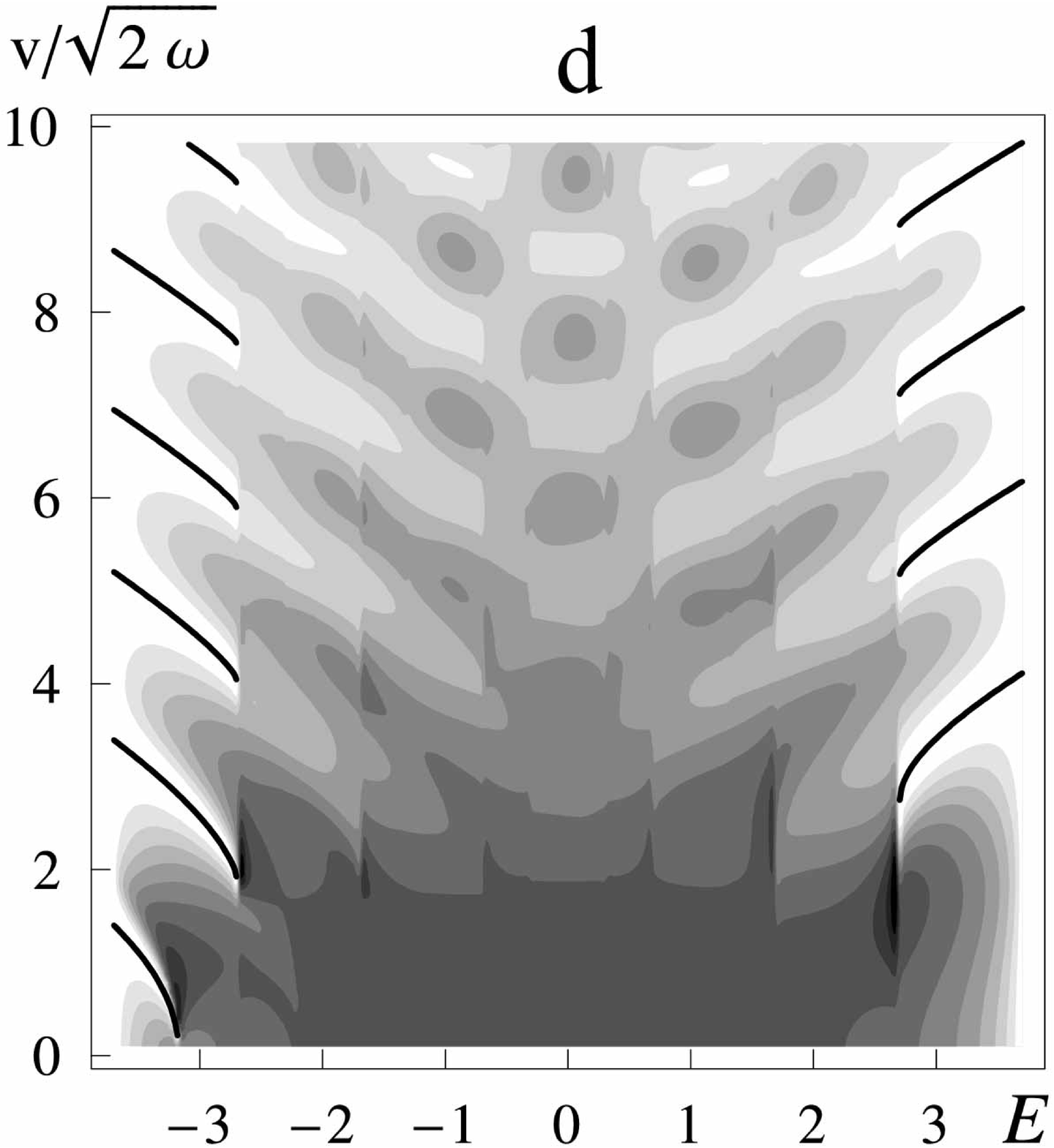}}\vspace{0.1cm}
\centerline{\epsfysize=3.7cm\epsfbox{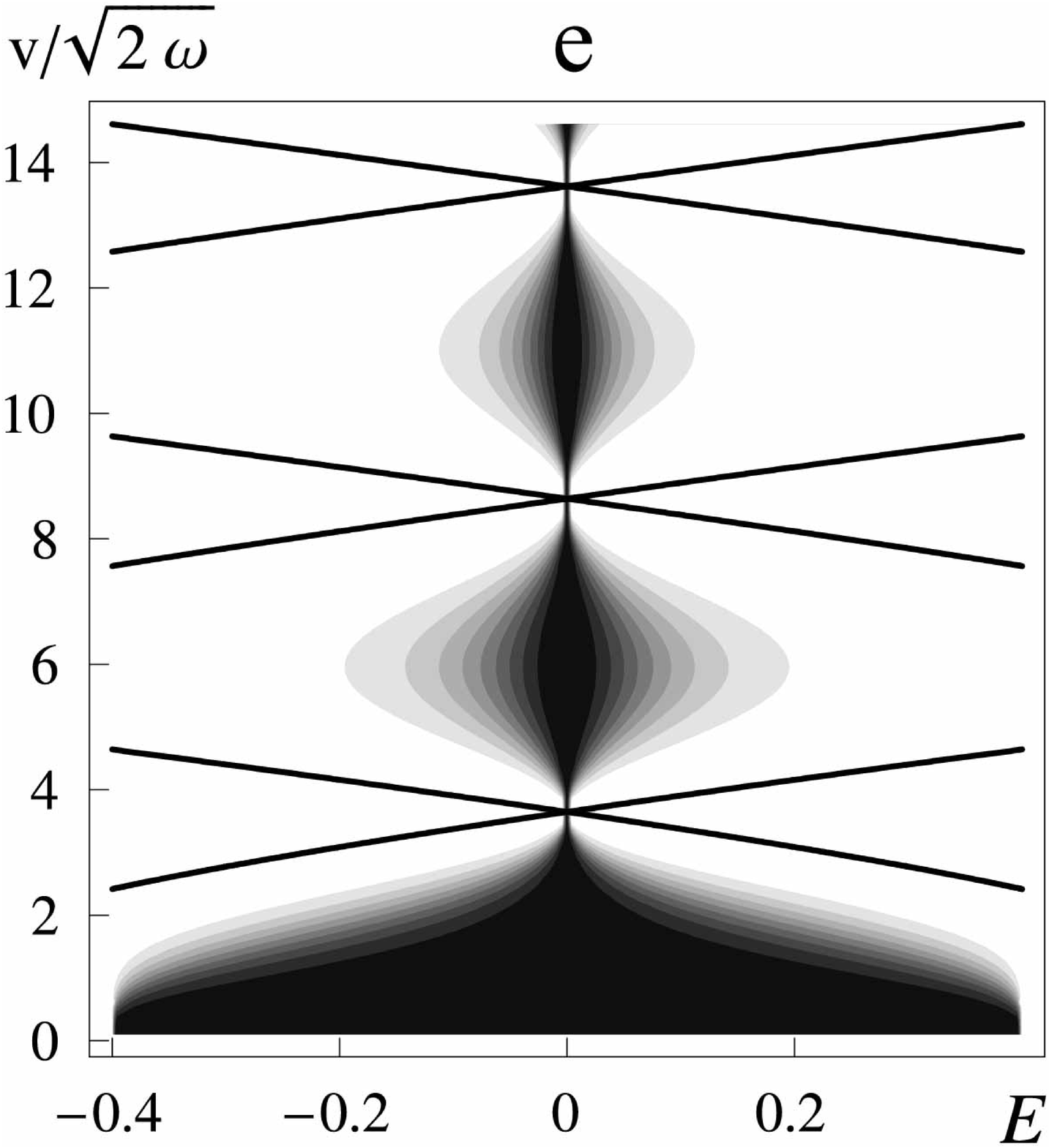}\hspace{0.2cm}\epsfysize=3.7cm\epsfbox{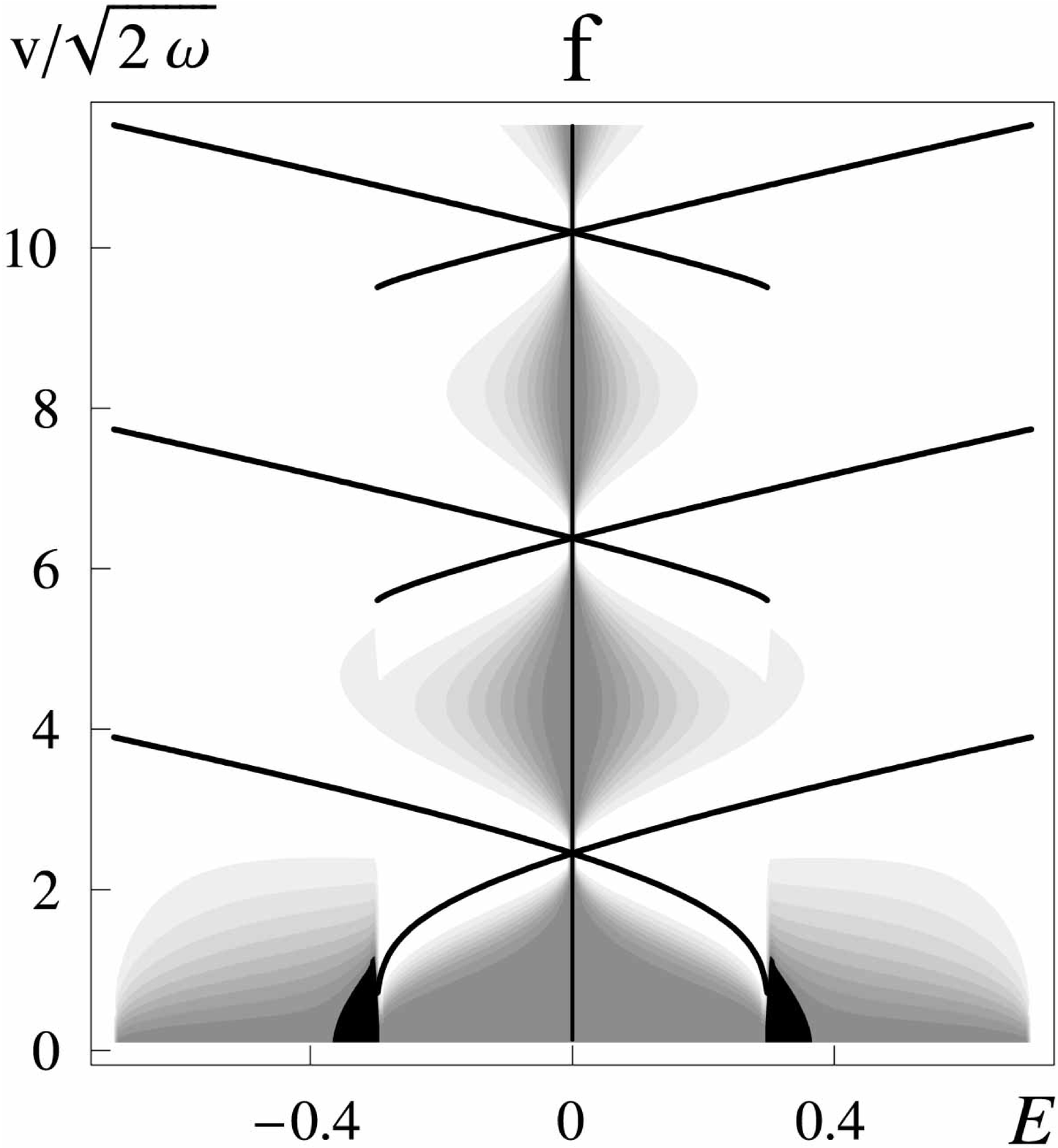}}\vspace{0.1cm}
\caption{(a, c-f) - Relief of conductance in the tight-binding
model as a function of the Fermi energy (in units of $\omega$) and
$v$. Figures a, c, d  with $\Delta=3.7$ differ by values of
$u=0$(a), 3(c), -1(d); Figs. e and f with $u=0$ differ by
$\Delta=0.4$(e), 0.7(f).  Levels of conductance change equidistant
from $<0.1$ (white) to $>1$ (black). Black lines represent the
blockade states. Vertical line $E=0$ corresponds to the
reflectionless state. b - Conductance versus $E_F$ for
$\Delta=3.7$, $u=0$; $v=\sqrt{2}$ (dash-dotted), $v=2\sqrt{2}$
(dashed), $v=3\sqrt{2}$ (dots), $v=4\sqrt{2}$ (line).}\label{map1}
\end{figure}

Figures~\ref{map1}a, \ref{map1}c-\ref{map1}f present conductance
relief as a function of $E_F$ and $v$. Sophisticated relief
reflects the photonic repetitions of the permitted band boundaries
(see Fig.~\ref{map1}b). The blockade states, situated in the
$\omega$ vicinity of the permitted band boundaries, are indicated
in the figures with the bold lines.

In addition to the blockade states, for $\Delta<\omega$ the
reflectionless state exists. If $u=0$, this state corresponds to
$E=0$. Its trace crosses the lines of blockade, forming
essentially singular points with directional dependent limits. The
case $u=0$ (Figs.~\ref{map1}a, \ref{map1}e, \ref{map1}f) gives
symmetric pictures with respect to the permitted band center
$E=0$, while the finite value of $u$ brings asymmetry
(Figs.~\ref{map1}c, \ref{map1}d).

Together with the considered discrete states, this model possesses
local states if $\Delta<\omega$ and $u\neq 0$. The existence of
these states is conditioned by the impossibility of fulfillment of
the energy conservation law for electron excitation from local
state to continuum. Such states are impossible in the model of
vibrating $\delta$-functions.

\vspace{0.1 cm}\noindent{\bf Conclusions.}\vspace{0.1 cm}

Thus, we have demonstrated that in different models with local
oscillating potential the states with zero transmission exist.
This fact is unusual, because such states are impossible for
static potential with limited integral. Together with these
blockade states we have found the reflectionless and local states
in the tight-binding model with oscillating level of a single
site. The presence of blockade results in vanishing of conductance
in considered systems. These systems can be used as quantum pumps.
In the blockade conditions for symmetric double-well systems the
derivative of photocurrent with respect to the Fermi energy also
vanishes, so that the photocurrent itself has step-like behavior.
So, the behavior of conductance and photocurrent looks like
behavior of $\sigma_{xx}$ and $\sigma_{xy}$ in quantum Hall
effect. Additionally, existence of blockade states leads in double
wells to possible strong localization of states lying in the
continuum.

We hope that the considered systems  can be realized and utilized
for confinement of electrons and fundamental measurements.


The work was supported by grant of RFBR No 05-02-16939, Program
for support of scientific schools of the Russian Federation No.
4500.2006.2, the grant of the President of the Russian Federation
No. MK-8112.2006.2 and the Dynasty Foundation.

\end{document}